\documentclass[useAMS]{mn2e}
\usepackage{epsfig,amssymb,amsmath}
\usepackage{subfigure}

\title{An expanded merger-tree description of cluster evolution}
\author[I. Dvorkin and Y. Rephaeli]{Irina Dvorkin$^{1}$\thanks{E-mail:
irina@wise.tau.ac.il} and Yoel Rephaeli$^{1,2}$\\
$^{1}$School of Physics and Astronomy, Tel Aviv University, Tel Aviv, 69978,
Israel\\
$^{2}$Center for Astrophysics and Space Sciences, University of California,
San Diego, La Jolla, CA 92093-0424}

\begin{document}

\pagerange{\pageref{firstpage}--\pageref{lastpage}} \pubyear{2010}

\maketitle

\label{firstpage}

\begin{abstract}
We model the formation and evolution of galaxy clusters in 
the framework of an extended dark matter halo merger-tree algorithm that
includes baryons and incorporates basic physical considerations. 
Our modified treatment is employed to calculate the probability density
functions of the halo concentration parameter, intracluster gas temperature, and
the integrated Comptonization parameter for different cluster masses and 
observation redshifts. Scaling relations between cluster mass and these 
observables are deduced that are somewhat different than previous results. 
Modeling uncertainties in the predicted probability density functions are 
estimated. Our treatment and the insight gained from the results presented 
in this paper can simplify the comparison of theoretical predictions with
results from ongoing and future cluster surveys.

\end{abstract}


\section{Introduction}

Formation of galaxy clusters is of central importance for understanding 
the evolution of the large scale structure (LSS) of the universe.
Statistical properties of clusters - deduced from cluster optical, 
X-ray, and SZ surveys - can be used to determine the
basic cosmological parameters - such as the matter density, the
normalization (and spectrum) of primordial density fluctuations, 
and the dark energy equation of state - independently of other methods
(CMB power spectrum, galaxy surveys, etc). 
This can only be achieved if theoretical tools are 
developed for a quantitative description of the non-linear hierarchical 
growth of clusters. Hydrodynamical numerical simulations of clusters 
are currently the most versatile tool for establishing statistical properties 
of clusters, which are commonly specified in terms of mass functions and 
scaling relations between their intrinsic properties.

In order to use cluster surveys as cosmological probes it is essential to 
know how their intrinsic properties - such as formation time, mass and 
redshift - affect their integrated statistical properties. 
Clusters are commonly described as virialized spherical systems whose masses 
are dominated by dark matter (DM) and with isothermal intracluster (IC) 
gas as their main baryonic mass component. Cluster statistical relations 
are usually expressed in terms of the mass and the redshift of observation; 
for example, the resulting scaling of the gas temperature is
\begin{equation}
	k_BT\propto M^{2/3}\left[E^2(z)\Delta_V(z)\right]^{1/3} ,
    \label{eq:Tscaling}
\end{equation}
where $E(z)=H(z)/H_0$, the Hubble parameter in units of its present
value and $\Delta_V(z)$ is the overdensity at virialization.

However, this simple description may not be sufficiently
adequate for the description of real clusters. Observations and numerical
simulations indicate that DM density profiles are shallower than isothermal at
small radii, and steeper than isothermal at large radii, and are characterized
by a scaling radius $r_s$ which marks the transition between these two regions 
(Navarro, Frenk \& White 1995, 1996). Although it seems reasonable to assume
that the scaling relations are not much affected by the details of the cluster
structure, the standard description is not sufficiently accurate since the model
parameters - such as $r_s$ - depend on the cluster mass and redshift. 
Indeed, numerical simulations show that the formation history of clusters 
affects the DM scaling radius, such that clusters are systematically 
denser the earlier they formed. Cluster mass concentration is usually 
quantified in terms of the parameter $c$, defined as the ratio of the 
cluster virial radius to the scaling radius, $c=R_v/r_s$. Thus, clusters 
that form earlier have larger concentration parameters (Wechsler et al. 
2002); more generally, other physical properties are also influenced by 
the cluster formation history.

The assumption of hydrostatic equilibrium is obviously an approximation,
since clusters, the largest bound systems, are still forming through
mergers of subclumps and accretion. Mergers disrupt the state of thermal
equilibrium; during some merger events IC gas temperature and X-ray
luminosity are boosted up by factors of up to $3$ and $10$, respectively,
as was demonstrated in a series of numerical simulations by Ricker and
Sarazin (2001). More generally, when deriving the standard scaling relations
one usually ignores the formation history of the cluster. In the $\Lambda$CDM
model the LSS formed hierarchically, through consecutive mergers of smaller
structures, and thus clusters with roughly the same mass and redshift may have
very different merger and formation histories. Indeed, both observations and
simulations reveal scatter in the mass-observable relations at a level of
$\sim 10\%-20\%$, which is partly due to the different formation histories 
(e.g., Wechsler et al. 2002, Vikhlinin et al. 2009a). In order to use these 
relations to determine the cluster mass function from surveys it is 
important not only to understand the exact scaling relations, but also to 
quantify the amount of scatter. Furthermore, uncertainties in the 
mass-observable relation reduce the precision with which cosmological 
parameters can be determined (Lima \& Hu 2005; Cunha \& Evrard 2009).

In current analyses of cluster surveys the scatter in the mass-observable 
relations has been partly accounted for. Some of this scatter is clearly due 
to the different dynamical state of the clusters in the survey; unrelaxed 
systems are sometimes excluded from the analysis. Several studies have 
indicated that the scatter in the scaling relations is reduced if IC gas 
temperature is measured by excluding the cluster central region (which is 
significantly affected by radiative cooling), but the intrinsic scatter in 
the scaling relations is more difficult to estimate. For example, motivated 
by insight from simulations, Vikhlinin et al. (2009b) used a constant value 
of $20\%$ as an estimate for the $T_X-M$ scatter. However, the amount of 
scatter may depend on the mass and redshift of the cluster.

A meaningful comparison with observational data usually necessitates 
knowledge of the full probability distribution function (PDF) rather than 
just the scaling with mass. An example is the case of very large values of 
the concentration parameter and Einstein radius for several clusters 
(Broadhurst et al. 2008, Zitrin et al. 2009). Based on results from N-body 
simulations, the probability of observing clusters with the
very high measured values was found to be very low, amounting to a
$4-\sigma$ discrepancy with the $\Lambda$CDM predictions. The
concentration parameter PDF is a crucial component in this analysis 
(Sadeh \& Rephaeli 2008). It is clear that a better understanding of the 
origin of this PDF is required.

The impact of the formation history on the cluster properties was previously 
investigated using a series of hydrodynamical simulations (Ricker and
Sarazin 2001, Randall and Sarazin 2002, Wik et al. 2008). These authors
used simulations of pairs of merging clusters and analysed the impact of
recent mergers on IC gas temperature, luminosity, and Comptonization
parameter. They found that all these quantities are boosted for a 
relatively long time following a merger, and calculated the effect of 
this boost on cosmological parameter estimation. These analyses did not 
include the full formation history, only the latest merger, and relied on 
a small number of simulated clusters with different masses. Voit and 
Donahue (1998) showed that the temperature evolves less rapidly with mass 
than in the standard analysis when the recent formation approximation is 
relaxed. They assumed gradual mass accretion throughout the cluster history, 
and a one-to-one correspondence between the temperature and the virial 
energy of the cluster.

The statistics of DM halo concentrations and their dependence on the halo 
formation history were investigated using N-body simulations (Bullock et al. 
2001, Wechsler et al. 2002, Neto et al. 2007, Gao et al. 2008, Duffy et al. 
2008). This approach provides PDFs of the concentration parameter which 
account for different formation histories of different halos. Results of 
these works can then be used to infer the intrinsic scatter in other cluster 
observables. However, a theoretical approach that is complementary to N-body 
simulations is needed in order to fully understand the impact of the cluster 
formation history on its properties. The reliability of the statistical 
analysis of N-body simulations depends on the simulation volume, which is 
limited by computational constraints. This can be a severe problem if one is 
interested in high-mass clusters, which are relatively rare systems. For 
example, the Millennium Simulation (MS, Springel et al. 2005), the largest 
cosmological N-body simulation to date, which follows $N=2160^3$ particles 
in a periodic box of $L=500h^{-1}$ Mpc on a side, contains less than $800$ 
relaxed halos with masses above $M_{200}=1.3\cdot10^{14}h^{-1}M_{\odot}$, and 
just $8$ relaxed halos with masses above 
$M_{200}=7.5\cdot10^{14}h^{-1}M_{\odot}$ (Neto et al. 2007, hereafter N07). 
Moreover, comparison between different simulations is difficult because each 
utilizes different cosmological parameters and halo finding algorithms. This 
difficulty is illustrated by the fact that somewhat different mass functions 
are predicted by different N-body simulations (Jenkins et al. 2001, Sheth \& 
Tormen 2002, Tinker et al. 2008), most likely reflecting the different 
formation histories predicted by these simulations.

Predicting the full PDFs of the relevant cluster parameters in the context 
of a theoretical model that can be readily implemented, would allow a more 
meaningful statistical analysis and the ability to quantify the impact of 
uncertainties in the values of cosmological and cluster parameters on the 
main observables deduced from large scale surveys. In this paper we develop a 
model of cluster formation using analytically computed DM merger trees, with 
which we trace the formation of clusters through major episodal mergers and 
continuous accretion. We show that our approach provides an improved physical 
description in comparison with what can be obtained from standard scaling 
relations.

This paper is organized as follows. In Section \ref{sec:method} we describe 
our model of cluster formation. The results of our generalized merger-tree 
treatment are presented in Section \ref{sec:results} and further discussed in 
Section \ref{sec:discussion}. Throughout the paper we use the following 
cosmological parameters: $\Omega_m=0.25$, $\Omega_{\Lambda}=0.75$, $H_0=73$ 
km/s/Mpc, $\sigma_8=0.8$. 

\section{Methodology}
\label{sec:method}

Our description of the growth of galaxy clusters is based on merger 
trees of DM halos as described numerically by the modified GALFORM 
code (Cole et al. 2000; Parkinson, Cole \& Helly 2008), which was
successfully employed to construct semi-analytic models of galaxy formation. The
algorithm implements the excursion set formalism, a key aspect of which is
the conditional mass function: the fraction of mass
$f(M_1|M_2)$ from halos of mass $M_2$ at redshift $z_2$ that
consisted of smaller halos of mass $M_1$ at an earlier redshift $z_1$
\begin{equation}
\begin{split}
	& f(M_1|M_2)d\ln M_1 = \\
	& \sqrt{\frac{2}{\pi}}\frac{\sigma_1^2(\delta_1-
\delta_2)}{(\sigma_1^2-\sigma_2^2)^{3/2}}\exp\left(-\frac{1}{2}\frac{(\delta_1-
\delta_2)^2}{(\sigma_1^2-\sigma_2^2)}\right) \frac{d\ln\sigma_1}{d\ln M_1}d\ln M_1 , 
\end{split}
\label{eq:cond_mf}
\end{equation}
where $\sigma_i=\sigma^2(M_i)$ is the variance of the linear 
perturbation field smoothed on scale $M_i$, and $\delta_i$ is the 
critical density for spherical collapse at redshift $z_i$. The 
original GALFORM algorithm is consistent with the Press-Schechter mass 
function (Press \& Schechter, 1974; PS), in the sense that if a grid of 
trees is rooted at $z=0$, weighted by their mass abundance according to 
the PS mass function, then the mass function at higher redshifts again 
corresponds to PS mass function. The conditional mass function is used 
to calculate the mean number of progenitors of mass $M_1$ at redshift 
$z_1+dz_1$ of a halo of mass $M_2$ at $z_2=z_1$:
\begin{equation}
	\frac{dN}{dM_1}=\frac{1}{M_1}\left(\frac{df}{dz_1}\right)_{z_1=
z_2}\frac{M_2}{M_1}dz_1
\label{eq:mean_numb}
\end{equation}
The modified GALFORM algorithm was obtained by making the substitution
\begin{equation}
	\frac{dN}{dM_1} \rightarrow \frac{dN}{dM_1}G(\sigma_1/\sigma_2,
\delta_2/\sigma_2) , 
\label{eq:G_function}
\end{equation}
where $G$ is referred to as a perturbing function, to be calibrated 
by comparison with N-body simulations. Parkinson et al. (2008) showed 
that by fitting the outcome of the algorithm to the results of the MS 
they obtained halo abundances which are consistent with the
Sheth-Tormen mass function (Sheth \& Tormen 2002). Thus, the perturbing function
$G$ expresses the uncertainty in the choice of the correct mass function. In
this work we use the following parametrization of the perturbing function:
$G=G_0\left(\sigma_1/\sigma_2\right)^{\gamma_1}
\left(\delta_2/\sigma_2\right)^{\gamma_2}$, with the parameters $G_0=0.57$,
$\gamma_1=0.38$, $\gamma_2=-0.01$ taken from Parkinson et al. (2008).

Starting with the specified mass and redshift, the algorithm proceeds 
back in time, checking after each timestep whether there was a merger; 
if so, the masses of the merged halos are drawn from the distribution 
(\ref{eq:mean_numb}). Halos with masses below some resolution limit 
$M_{res}$ are not resolved, and are accounted for as continuously 
accreted mass. Further details on the GALFORM algorithm can be found 
in Parkinson et al. (2008).

In the following sections we describe our model of cluster formation. 
Since the DM is the dominant mass component of clusters we first study 
the formation history of cluster-sized DM halos, and then add the IC 
gas component, study its properties and related scaling relations.

\subsection{Modeling the formation of dark matter halos}
\label{sec:dm_model}

The calculation begins with the construction of a merger tree for a given final 
halo mass and redshift. For each tree, we use only the major merger events, that 
is only those mergers for which $M_{>}/M_{<}< q$, whose value is to be determined. 
The rationale behind this choice is the assumption that the properties of a cluster 
are largely determined by the violent merger events, during which DM and gas settle 
in the modified potential well when equilibrium is reestablished. On the other 
hand, slow accretion of material on the outskirts of the cluster, as well as 
mergers with low-mass systems (which are treated in an identical manner in this 
work) do not cause redistribution of the cluster components and do not strongly 
affect the physical conditions in the cluster center. Thus, we follow the major 
merger events in the cluster history, and refer to all other processes of mass 
growth as slow accretion. A typical value of the major merger ratio is $q=10$; 
other values will also be considered. The trees were generated up to the redshift 
$z=z_{fin}$ which depends on the mass resolution of the merger tree algorithm, 
$M_{res}$, as discussed below.

The next step is to calculate the density profile of each halo in the tree. To this 
end, the tree is traced down, beginning with the smaller masses; for each merger 
event we use energy conservation to calculate the density profile of the merged DM 
halo. For simplicity we assume NFW (Navarro, Frenk \& White 1995) profiles for all
the halos at all times:
\begin{equation}
	\rho_{DM}(r)=\frac{\rho_s}{x(1+x)^2}
	\label{eq:nfw}
\end{equation}
where $x=r/r_s=rc/R_v$ is the radial distance expressed in terms of NFW scale 
radius $r_s$, $c$ is the concentration parameter, and $\rho_s$ is the mass 
density normalization constant. Each halo is completely characterized by its 
mass, redshift and concentration parameter, where halo mass is defined as the 
mass within the virial radius, $M=\frac{4\pi}{3}R_v^3\rho_c\Delta_V$. All 
masses are assumed to be in equilibrium, which is presumed to be attained 
relatively quickly after each merger event, but we do not assume the halos 
are completely virialized within the virial radius. In fact, the virial ratio 
$2T/|W|$ approaches $1$ at the virial radius only for very large $c$, while for 
commonly deduced values of 
$c$ this ratio is slightly larger than $1$ at the virial radius (Cole \& Lacey 
1996; {\L}okas \& Mamon 2001).

The relation between mass and virial radius depends on redshift, which we identify
as the halo formation redshift, $z_f$, defined as follows. If the halo has undergone 
a major merger event, we take $z_f$ to be the redshift of the last major merger.
This choice is consistent with the main assumption that major mergers largely 
determine the physics of the halo. For halos that did not experience major mergers 
at all, and were, according to our interpretation, entirely assembled by 
minor mergers and continuous
accretion, we take $z_f$ to be the redshift at
which half of the halo mass had assembled. Difficulty in determining $z_f$ by this 
prescription is encountered only when the branch of the merger tree terminates when 
the halo still has \emph{more} than half of its mass (recall that the tree is 
evolved backwards in time). In GALFORM, the branch is terminated in two cases: 
either $z_{fin}$ is reached, or the mass of the halo falls below the resolution 
mass $M_{res}$. Thus, in order to describe the formation of the smallest halos that 
constitute the tree, $z_{fin}$ should clearly be chosen well above the expected 
formation redshift of halos of mass $M_{res}$. By making this choice we ensure that 
almost all of the halos in the tree assemble half of their mass before $z_{fin}$ 
is reached (that is, later in time), and their formation redshift can be traced 
back by the tree.

To start from the earliest halos in the tree and move forward in time requires 
specifying their concentration parameters. These are adapted from a fit to a set 
of N-body simulations by Bullock et al. (2001) 
\begin{equation}
	c(M,z_{obs})=5\left(\frac{M}{10^{15}h^{-1}M_{\odot}}\right)^
{-0.13}\frac{(1+z_f)}{(1+z_{obs})} , 
\label{eq:cfit_bullock}
\end{equation}
where $z_f$ and $z_{obs}$
are the redshifts of formation and observation, respectively. 
This choice is motivated by the finding of Wechsler et al. (2002) that the 
concentration parameter scales as $c\sim (1+z_f)/(1+z_{obs})$, although note that 
their definition of the formation redshift is slightly different than the one 
adopted here. Although this choice for the initial $c(M,z)$ is somewhat arbitrary, 
its particular form does not significantly influence the results.

For each merger event, we calculate the total energy of the system before merging, 
which depends on the concentration parameters of the merging halos, and deduce the 
concentration parameter of the merged halo from simple energy conservation 
arguments, motivated by a cluster merger model by Sarazin (2002). The total 
energy of a system of two halos before merging is:
\begin{equation}
	E_{tot,12}=E(M_1)+E(M_2)+U_{12}
\end{equation}
where $E(M_i)$ are the total energies (potential and kinetic) of each halo and 
$U_{12}$ is the gravitational energy of the two halos at the point of their 
largest separation, when they have just become bound and
their relative velocity was negligible:

\begin{equation}
	U_{12}=-\frac{GM_1M_2}{d_0}
\end{equation}

The distance $d_0$, at which the halos became bound, is roughly the mean distance 
between halos with masses $M_1$ and $M_2$ that reside in an overdense region with a 
scale that corresponds to the final mass $M_f=M_1+M_2$. We take this distance to be
$d_0=\kappa d$, where
$d=R_1+R_2$, and adopt $\kappa=5$ as a fiducial value for all halos.
This corresponds to a typical distance of several Mpc and a typical relative 
velocity of several hundreds to a few thousands of km/s, depending on the masses 
of the merging clusters, which is in accordance with the initial conditions of 
hydrodynamical cluster merger simulations (Ricker \& Sarazin 2001, McCarthy et 
al. 2007, Lee and Komatsu 2010). Very high values of $\kappa$ produce 
unrealistically large initial separations and large relative velocities, while 
very low values of $\kappa$ lead to very small initial distances, small relative 
velocities, and consequently, greatly reduced total energies, which eventually 
result in very high concentration parameters of the final halo. In other 
words, $\kappa$ was chosen so as to yield realistic values of both the initial 
separation and the final concentration parameter. The dependence of the results on 
$\kappa$ is discussed below.

After the two halos merge, the resultant halo accretes matter, so when in turn it
merges to form a larger halo, it has more mass than the sum of the masses of its 
progenitors. We account for the energy of the accreted matter in a very approximate 
way as follows. Given the masses of the two progenitor halos, $M_1$ and $M_2$, and 
the final mass of the halo, $M_p$, the total accreted mass is 
$\Delta M=M_p-(M_1+M_2)=M_p-M_f$. 
Numerical simulations of galaxy-sized halos (e. g. Wang et al. 2010) seem
to indicate that the accreted material is distributed in the halo outer
region. This is quite likely the case also in cluster-sized halos, so we
can estimate the energy due to the accreted mass by writing
\begin{equation}
	U_{acc}=-\frac{G(M_1+M_2)\Delta M}{R_f} ,
	\label{eq:uacc}
\end{equation}
where $R_f$ is the virial radius of the halo with mass $M_f=M_1+M_2$ that formed
just after the merger. In evaluating $U_{acc}$ we assume that the accreted mass 
constitutes a relatively small fraction of the final halo, and that equation 
(\ref{eq:uacc}) provides a simplified description of the accretion process.

We then have for the total energy of the system prior to merging
\begin{equation}
	E_{total}=E_{tot,12}+U_{acc}
\end{equation}
This energy is attributed to the resulting merged halo; we assume no mass is
lost in the process. By equating $E_{total}$ with the gravitational and
potential energy of the resulting halo, which depend on its concentration
parameter, we can deduce the latter. This process is repeated for each halo in
the tree, until arriving at the bottom - the most massive halo. At the end of 
this process we obtain the concentration parameter for the given mass and for 
one tree, i.e. one realization of the halo history. Generating a large number 
of trees gives an estimate of the PDF of $c(M)$.

\subsection{Modeling the intracluster gas}
\label{sec:gas_model}

In our modeling of IC gas we assume that it constitutes a small fraction of the
total cluster mass, and that it does not significantly affect the evolution of 
the cluster. We assume that the gas has a polytropic equation of state with 
an adiabatic index $\Gamma$, such
that the gas density and pressure are related through:
\begin{equation}
	P=P_0(\rho/\rho_0)^{\Gamma}
\end{equation}

The solution of the equation of hydrostatic equilibrium for a polytropic gas
inside a potential well of a DM halo with an NFW profile is
(Ostriker, Bode, \& Babul 2005):
\begin{equation}
	\rho(x)=\rho_0\left[1-\frac{B}{1+n}\left(1-\frac{\ln(1+x)}{x} \right)
\right]^{n} ,
\end{equation}
where $n=(\Gamma-1)^{-1}$ and $B$ is given by:
\begin{equation}
	B=\frac{4\pi G\rho_s r_s^2\mu m_p}{k_B T_0} , 
\end{equation}
and $\mu m_p$ is the mean molecular weight. Thus, $B$ depends on the
concentration 
parameter through $\rho_s$ and $r_s$ (see equation (\ref{eq:nfw})).
The temperature profile is then
\begin{equation}
	T(x)=T_0\left[1-\frac{B}{1+n}\left(1-\frac{\ln(1+x)}{x} \right) \right]
\end{equation}

As a boundary condition we assume that the gas pressure at the virial radius
obeys $P_{gas}=f_gP_{DM}$ where $f_g$ is the gas mass fraction and
$P_{DM}=\rho_{DM}\sigma^2$ (Ostriker et al. 2005). We obtain $\sigma^2$, the DM
(3D) velocity dispersion, by solving the Jeans equation for the NFW potential.
We expect this particular choice for the boundary condition to have a minor
influence on the results, as discussed in Section \ref{sec:discussion}.

For each merger tree we obtain the concentration parameter and the
virial radius of the final halo, as described above.
Taking the observationally deduced value for the adiabatic index,
$\Gamma=1.2$, and imposing the above boundary condition 
to obtain the constant $B$ fully determines the temperature profile. By assuming
a specific gas mass fraction we also obtain the full density profile. Repeating
this
procedure for a large number of trees provides an estimate of the PDFs 
of the various physical parameters as a function of the cluster mass and 
the redshift of observation.

In order to derive scaling relations we define the mean cluster temperature as
the
emission-weighted value
\begin{equation}
	T_{ew}\equiv \frac{\int{\Lambda(T)\rho_g^2T dV}}{\int{\Lambda(T)\rho_g^2
dV}} , 
	\label{eq:Tew}
\end{equation}
where integration is over the cluster volume, and the temperature dependence of
the cooling function is approximately $\Lambda(T)\propto \sqrt{T}$. Similarly,
X-ray luminosity is approximated by:
\begin{equation}
 L_{X}=\int{\left(\frac{\rho_g}{\mu m_p}\right)^2 \Lambda(T)dV} .
\end{equation}

Knowledge of the gas temperature allows also calculation of the Comptonization 
parameter, $y$, defined as 
\begin{equation}
	y=\int{\left(\frac{k_BT_e}{m_ec^2} \right)n_e\sigma_T dl} ,
	\label{eq:SZ_y}
\end{equation}
where the integral is taken along the line of sight to the cluster, $\sigma_T$
is 
the Thomson scattering cross section (and the electron temperature $T_e$ is
assumed 
to equal the gas temperature).
The measured SZ intensity (change)
is approximately proportional to the integrated Y-parameter, given by 
the integral of $y$ over the angle the cluster subtends on the sky
\begin{equation}
	Y=\int{yd\Omega}
	\label{eq:SZ_Y}
\end{equation}

\section{Results}
\label{sec:results}

\subsection{Model parameters}

The model contains several
physical parameters, and two
numerical (code-specific) parameters - the number of tree realizations
used to estimate the PDFs and the resolution mass of the merger tree.
We shall discuss the latter parameters here and defer the discussion of the 
physical parameters to subsequent sections.

A key objective of our model
is to determine the PDF of the concentration parameter and IC gas temperature 
by generating a large number of merger tree realizations $N$. We have found 
that taking $N=5\cdot 10^4$ is sufficient to obtain convergent results -
taking larger $N$ does not change the PDF by more than a fraction of
a percent. In what follows, we show histograms of $50$ binned values obtained 
from $5\cdot 10^4$ merger trees.

As noted earlier, a resolution mass needs to be selected for each tree.
This mass is the smallest building block used in the tree. By
sampling different values of the resolution mass, we find that $M_{res}$ has to
be at least $3$ orders of magnitude below $M$, the final mass for which 
the tree is built, while taking smaller values of $M_{res}$ does not 
affect the results: for the mass range we consider, the mean value of 
$c$ changes by no more than $1\%$ when the resolution mass is lowered 
from $M_{res}=10^{-3}M$ to $M_{res}=10^{-4}M$. The value of $M_{res}$ 
determines $z_{fin}$, as discussed above.

\subsection{PDF of halo concentration}
\label{sec:c_dist}

\begin{figure}
\centering
\epsfig{file=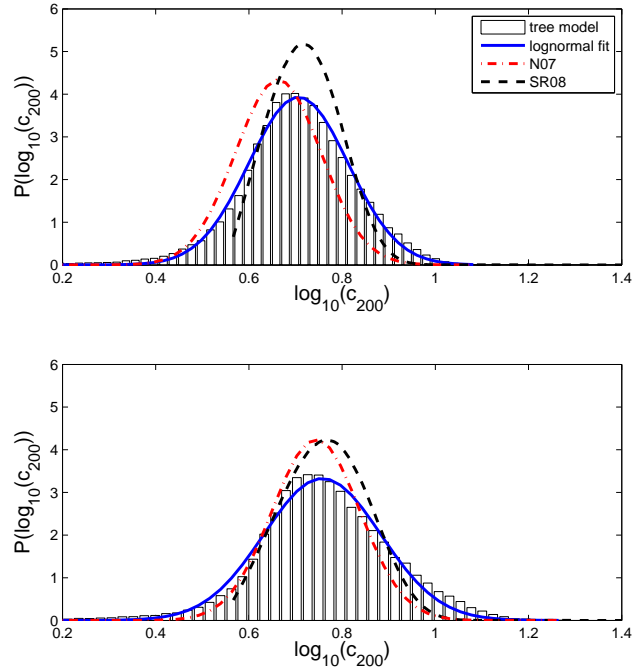, height=10cm}
\caption{Concentration parameter probability distribution function (PDF) as
predicted by the merger-tree model (bars), the lognormal fit to this PDF (solid
line), a distribution extracted from the MS by N07 (dot-dashed line) and the
prediction of SR08 (dashed line). \emph{Upper panel:} $M=4\times
10^{14}h^{-1}M_{\odot}$ for the merger tree model and a corresponding range of
$M_{200}=10^{14.25}-10^{14.75}h^{-1}M_{\odot}$ from the MS. \emph{Lower panel:}
$M=6\times 10^{13}h^{-1}M_{\odot}$ for the merger tree model and a corresponding
range of $M_{200}=10^{13.63}-10^{13.88}h^{-1}M_{\odot}$ from the MS.}
\label{fig:cc_dist}
\end{figure}

The basic outcome of the model is the concentration parameter of the DM 
halo at a given redshift of observation. Each merger tree results in a 
slightly different concentration parameter, which depends on the 
particular structure of the merger tree. Thus, in the limit of a large 
number of tree realizations the distribution of formation histories 
provides a PDF of the concentration parameter. 
The PDF of $c$ for $M=4\times 10^{14}h^{-1}M_{\odot}$ at $z=0$ is
shown in Figure \ref{fig:cc_dist} (upper panel).
A log-normal distribution provides a reasonable fit, with
$<\log_{10}c>=0.706$ and $\sigma_{\log_{10}c}=0.106$. The width of the
distribution is comparable with the value obtained by N07 for a
population of relaxed halos in the corresponding mass range
$M_{200}=10^{14.25}-10^{14.75}h^{-1}M_{\odot}$ seen in the 
MS: $<\log_{10}c>=0.663$ and $\sigma_{\log_{10}c}=0.092$. The distribution for a
lower mass of $M=6\times 10^{13}h^{-1}M_{\odot}$ is shown in the lower panel,
along with a corresponding distribution for halos in the MS in the mass range of
$M_{200}=10^{13.63}-10^{13.88}h^{-1}M_{\odot}$. For this mass we obtain
$<\log_{10}c>=0.758$ and $\sigma_{\log_{10}c}=0.106$ from the log-normal fit,
compared with $<\log_{10}c>=0.744$ and $\sigma_{\log_{10}c}=0.094$ for the
halos in the MS. We note that the mass correspondence is only approximate,
since the relation between $M_{200}$ and $M_v$ for a given halo depends on its
concentration parameter.

\begin{figure*}
\centering
\epsfig{file=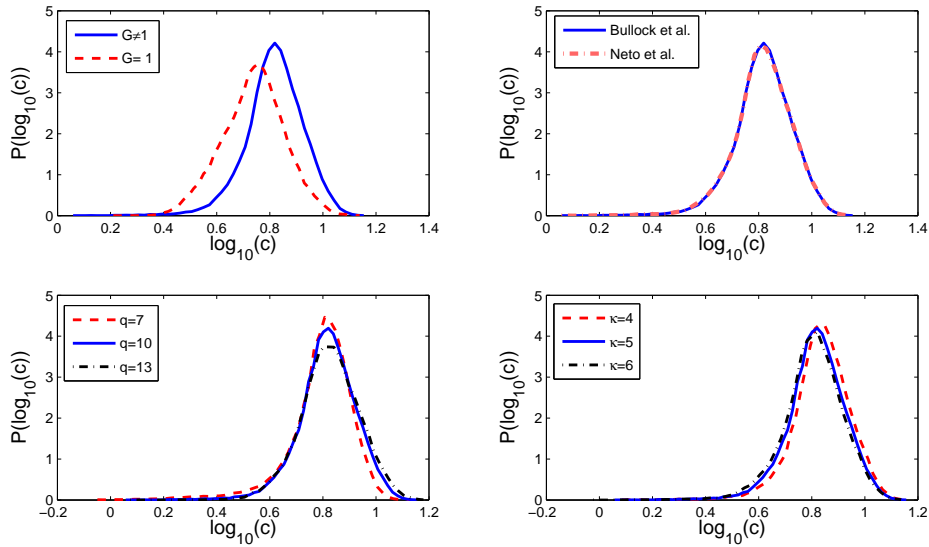, height=8cm}
\caption{PDF of the concentration parameter from the merger-tree model for
$M=10^{15}h^{-1}M_{\odot}$ at $z=0$ with different model parameters, as
discussed in the text. Shown in the \emph{Left upper panel} is the dependence on
the mass function through the perturbing function $G$ (see equation
(\ref{eq:G_function})), the dependence on the initial conditions for $c(M,z)$
for the earliest halos in the tree in the \emph{Right upper panel}, the
dependence on the major merger parameter $q=M_{>}/M_{<}$ in the \emph{Left lower
panel}, and the dependence on $\kappa=d_0/d$ (which parametrizes the initial
distance between halos that are about to merge) in the \emph{Right lower
panel}.}
\label{fig:cc_quadruple}
\end{figure*}

For comparison, a semi-analytical calculation adapted from Sadeh and Rephaeli 
(2008; SR08) is also shown. This latter treatment was based on an analytical 
distribution of formation times and a relation between the formation time and 
concentration parameter deduced from numerical simulations by Wechsler et al. 
(2002). The merger tree model predicts slightly lower concentration parameters
and a slightly broader distribution function. It is important to note that both
treatments result in quite similar PDFs that are also consistent with the
results of numerical simulations, despite of the competely different
assumptions made in each of these approaches.

As mentioned earlier, the uncertainty in the correct form of the mass function is 
quantified by the perturbing function $G$ (see equation (\ref{eq:G_function})).
Figure \ref{fig:cc_quadruple} (left upper panel) shows how the concentration
parameter changes when the merger tree is computed with and without the
perturbing function $G$. As expected, the concentration parameter tends to be
larger in the former case, reflecting the earlier formation time of halos in the
MS as compared with the extended Press-Schechter formalism (Wechsler et al.
2002). This result illustrates the rather strong dependence of the PDF of $c$ on
the mass function. This dependence has to be accounted for when comparing
results from observations and numerical simulations.

The initial conditions of the tree are the concentration parameters of 
the earliest halos. As indicated earlier, the particular choice of 
$c(M,z)$ for the earliest halos does not 
appreciably affect the final value of $c$,
as long as this choice is reasonable. For example, 
Figure \ref{fig:cc_quadruple} (right upper panel) shows the probability
distributions of 
$c$ with the initial $c(M,z)$ taken from the fit in equation 
(\ref{eq:cfit_bullock}), and a different fit adapted from the results 
of N07:

\begin{equation}
	c=5.26\left(\frac{M}{10^{14}h^{-1}M_{\odot}}\right)^{-0.1}\frac{1+z_f}{1+z_{obs}}
\end{equation}
It can be seen that there is no significant change in the distribution 
function. The influence of these initial condition on the results is 
further discussed at the end of section \ref{sec:c_scaling}.

The structure of the tree, and hence the calculation of $c$,
depends somewhat on the chosen ratio for major mergers, 
$q=M_{>}/M_{<}$. This dependence is shown in Figure 
\ref{fig:cc_quadruple} (left lower panel); the choice of $q$ is guided by
several physical considerations. On the one hand, it should not be too small,
because this would take into account only nearly equal-mass mergers.
Hydrodynamical simulations (Wik et al. 2008, McCarthy et al. 2007) show that
mergers with mass ratios as high as $q=10$ still lead to strong disruption of
equilibrium in the inner cluster region, and would thus need to be treated as
major merger events in our approach. The dynamical impact of taking higher
values of $q$ has not been explicitly explored in hydrodynamical simulations.
Accordingly, we selected this value to be the highest value of $q$ above 
which the mergers are approximated as continuous mass accretion.

The value of $\kappa$, which determines the separation at which two 
halos become bound, also influences the results quite appreciably. 
Figure \ref{fig:cc_quadruple} (right lower panel) shows that deviations from the
fiducial value of $\kappa=5$ can shift the distribution of $c$ due to
changes in cluster initial energies. As discussed earlier, the value 
$\kappa=5$ was chosen so as to produce realistic distances between 
clusters that are about to merge, and is consistent with estimates of 
relative velocities of merging clusters (Lee and Komatsu 2010).

\subsection{Scaling relations of the concentration parameter}
\label{sec:c_scaling}

\begin{figure}
\centering
\epsfig{file=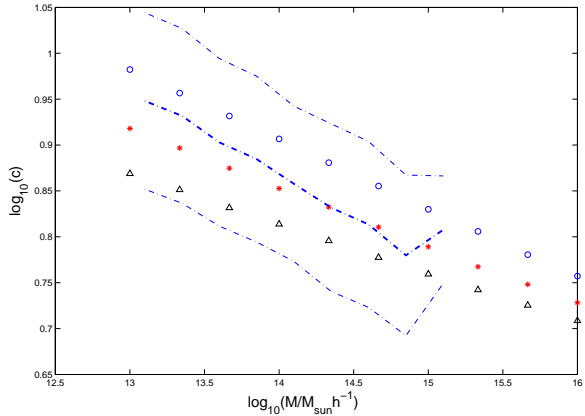,height=6cm}
\caption{$c-M$ relation at $z=0$ (circles), $0.5$ (stars) \& $1$
(triangles). The thick line shows the results of N07 with the
dispersion in the logarithm of $c$ indicated by the thin lines. The mass
dependence is consistent with the results of the MS.}
\label{fig:cmz_single}
\end{figure}

The expectation values of the distribution functions from the previous 
sections provide the concentration parameter averaged over formation 
histories. It is obviously important to follow the redshift evolution of $c$ and
its distribution with the final mass. Figure \ref{fig:cmz_single} shows the
$c-M$ relation for several redshifts of observation. The results can be well
described by the scaling relation $c\propto M^{-\alpha}$, with strong redshift
dependence of $\alpha$, ranging from $\alpha=0.075$ for $z=0$ to $\alpha=0.054$
for $z=1$, so that the dependence of $c$ on mass is weaker for higher redshifts.
This likely represents the fact that $c$ depends on mass through the formation
redshift, and the difference in formation redshifts for 
different masses observed at $z=0$ is larger than for different 
masses observed at a higher redshift. This flattening of the 
mass-concentration relation at high redshifts is also seen in numerical 
simulations (Duffy et al. 2008, Gao et al. 2008), although our predictions for
$\alpha$ are slightly lower at low redshift and slightly higher at high
redshift than those of Duffy et al. Figure \ref{fig:cmz_single} also shows the
results of N07 for halos at $z=0$ extracted from the MS (thick line)
along with the $1-\sigma$ distribution widths (thin lines). These results of the
MS are consistent with the predictions of the merger-tree model, although there 
seems to be a systematic offset between the 
respective results from these two very different studies.

The dependence on the cluster observation redshift, which is often taken to be
$c\sim (1+z_{obs})^{-\gamma}$ with $\gamma=1$, is also found to be much weaker
and mass-dependent, ranging from $\gamma=0.38$ for $M=10^{13}h^{-1}M_{\odot}$
to $\gamma=0.24$ for $M=10^{15}h^{-1}M_{\odot}$. This results in slower redshift
evolution than found by Duffy et al, but is more consistent with the findings
of Gao et al. for massive halos extracted from the MS, especially for masses
around $M\sim 10^{14}M_{\odot}$ for which our result $\gamma=0.31$ coincides
with the evolution seen by Gao et al. (note, however, that these authors use
the Einasto profile to describe DM halos). 

In general, the scaling relations deduced from numerical simulations are
effectively weighted by the mass function, and, since the latter has a sharp
cutoff at about the typical mass of a galaxy cluster, mainly reflect the
structure of smaller, galaxy-sized halos at low redshifts. Extrapolations of the
results of such simulations cannot faithfully describe the structure of massive
halos at high redshifts, as pointed out by Gao et al. Although we use the
results of Bullock et al. as the initial conditions for the merger tree -
equation (\ref{eq:cfit_bullock}) - this choice is justified because our final
results are not sensitive to the exact form of these initial conditions. In
addition, the initial halos in the merger tree have smaller masses, in the range
explored by Bullock et al.

Full investigation of the $c-M$ relations and their redshift evolution
neccesitates the use of numerical simulations targeted at massive,
cluster-sized halos. We plan to continue our study in this direction using the
hydrodynamical AMR code \emph{Enzo}.

\subsection{PDF of IC gas temperature}
\label{sec:temp}

\begin{figure}
\centering
\epsfig{file=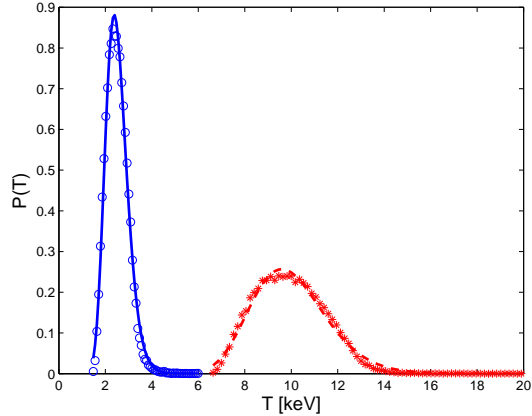, height=6cm}
\caption{IC gas temperature PDFs predicted by the merger-tree model and best 
fits to a log-normal distribution for $10^{14} h^{-1}M_{\odot}$ (circles, solid 
line), and for $10^{15} h^{-1}M_{\odot}$ (asterisks, dashed line). These 
distributions exhibit a characteristic sharp cutoff at low temperatures and a long 
high-temperature tail, as expected.} 
\label{fig:unifunc_m1415}
\end{figure}

Since the temperature of IC gas is used as a mass proxy in cluster 
surveys, its PDF is of great observational importance. We have 
computed this distribution as outlined above. Figure 
\ref{fig:unifunc_m1415} shows the PDFs of the emission-weighted
temperature for cluster masses $M=10^{15}h^{-1}M_{\odot}$ and
$M=10^{14}h^{-1}M_{\odot}$ at $z=0$.

As expected, the PDF exhibits a long high-temperature tail which 
corresponds to those clusters that were formed atypically early. 
At low temperatures, on the other hand, there is a sharp cutoff 
that corresponds to clusters that formed close to their observation 
redshift. A log-normal distribution provides a good approximation to the
temperature PDF below $M\sim 2\cdot 10^{15}h^{-1}M_{\odot}$, as can be seen in
Figure \ref{fig:unifunc_m1415}. The width of the distribution is
$\sigma_{\log_{10}c}=0.07$ for $10^{15} h^{-1}M_{\odot}$ and
$\sigma_{\log_{10}c}=0.08$ for $10^{14} h^{-1}M_{\odot}$.

\subsection{Temperature scaling relations}
\label{sec:tem_scale}

\begin{figure}
\centering
\epsfig{file=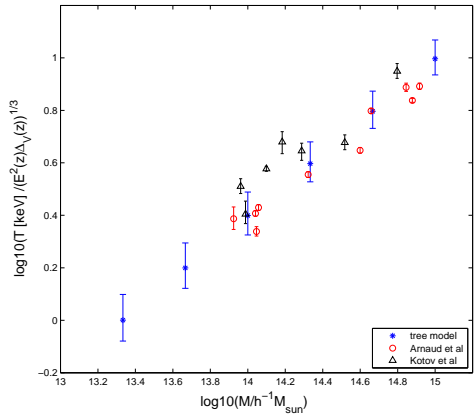, height=6cm}
\caption{Mass-temperature scaling relation at $z=0$ for the merger-tree model
(stars) and from observations (circles and triangles).}
\label{fig:TM_z0}
\end{figure}

Scaling relations of the gas temperature with cluster redshift, mass,
and X-ray luminosity are commonly used in statistical analyses of the 
cluster population and in the use of clusters as cosmological probes. 
Most useful is the $T-M$ relation which can be determined
from the probability distribution functions. In Figure \ref{fig:TM_z0}
we show the emission-weighted temperature versus mass for clusters at 
$z=0$. The temperature was calculated using equation (\ref{eq:Tew}). 
Blue stars represent expectation values of the PDFs, with errorbars indicating
the distribution variance. The red circles are measurements of a sample of
clusters from Arnaud, Pointecouteau \& Pratt (2005), and the black triangles are
measurements of another 
sample by Kotov \& Vikhlinin (2005), where redshift correcting
factors have been included for both samples. The merger tree results 
are best-fit with the relation $T\propto M^{0.6}$, which is very close to
the theoretical relation obtained for an isothermal sphere, 
$T\propto M^{2/3}$. It can be seen that the results and the expected
scatter are consistent with 
observations. Note though
the different definitions of mass ($M_{200}$ in Arnaud et al., $M_{500}$ 
in Kotov \& Vikhlinin) and temperature (spectral temperature in both Arnaud et
al. and Kotov \& Vikhlinin).

The observational results suggest that the variance of the temperature PDF
can be seen to represent the amount of scatter that is expected in 
observed clusters due to their different formation history. Note that 
the error in the measured temperatures is small compared to the 
scatter, which is slightly larger than the predicted intrinsic scatter, as
expected, since it has additional contributions. For example, not all clusters 
are fully relaxed and spherical, etc. We find that the temperature 
scales as a power-law in mass at all redshifts, $T\propto M^{\zeta}$; 
however $\zeta$ varies somewhat with redshift, from $0.6$ for 
$z=0$ to $0.63$ for $z=2$, which results in slower evolution compared 
to the simple scaling $\zeta=2/3$. We note that the \textit{minimal} 
possible temperature of a given mass - the low-temperature endpoint of 
our PDF - scales as $T_{min}\propto M^{0.65-0.66}$ in our model for all 
redshifts, in much better agreement with the standard value. Indeed, the 
standard treatment assumes that the halo is observed immediately after it 
had formed, which is precisely the situation described by the
low-temperature end of the PDF. The expectation value, however, is 
affected by the width of the PDF, which also depends on mass.

The predicted redshift dependence of $T$ is another key relation whose 
knowledge is important as it reflects on cluster evolution, and its approximate
analytic form is needed in comparisons with results of cluster X-ray and SZ
surveys. The basic redshift scaling of the temperature is contained in the
relation $T\propto [E^2(z)\Delta_V(z)]^{\lambda}$, where $\lambda$ varies 
somewhat with mass, from $\lambda=0.2$ for $M=10^{13}h^{-1}M_{\odot}$ 
to $\lambda=0.26$ for $M=10^{15}h^{-1}M_{\odot}$. Thus, $T(z)$ is less 
steep than in the standard relation (\ref{eq:Tscaling}), where 
$\lambda=1/3$. In addition, the slope of this scaling relation differs 
with mass, hinting that the temperature might not be a separable 
function in terms of mass and redshift. The dependence of these 
results on the model parameters is discussed in Section 
\ref{sec:model_unc}.

\begin{figure}
\centering
\epsfig{file=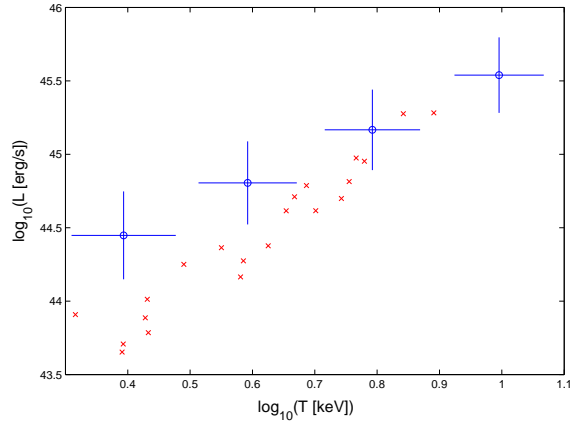, height=6cm}
\caption{X-ray luminosity vs. emission-weighted temperature (circles with error
bars), where the error bars correspond to the disrtibution
width in the logarithm of the luminosity and temperature, respectively. A
constant gas mass fraction $f_g=0.1$ was assumed. Also shown are measurements
of the bolometric luminosity vs. spectroscopic temperature from Pratt
et al. (crosses).}
\label{fig:LTp1}
\end{figure}

The luminosity-temperature relation is an important probe of the IC gas. In the
framework of the presented approach it can be used to test the validity of the
simple polytropic model. Figure \ref{fig:LTp1} shows the luminosity-temperature
relation obtained from the merger-tree model, as well as X-ray measurements of
a sample of clusters by Pratt et al. (2009). There is reasonable agreement with
the data in the high-temperature end, with the distribution width approximately
corresponding to the scatter in the measured values, but the model clearly
overpredicts the luminosity of low-temperature clusters. One reason for this
could be non-constant gas mass fraction which, as hinted by observations, is
lower in low-mass systems. The dependence of the gas mass fraction on mass 
and redshift could also be related to additional physical processes in the IC
gas, such as radiative cooling and feedback from supernovae and AGN.

\subsection{Integrated Comptonization parameter}

\begin{figure}
\centering
\epsfig{file=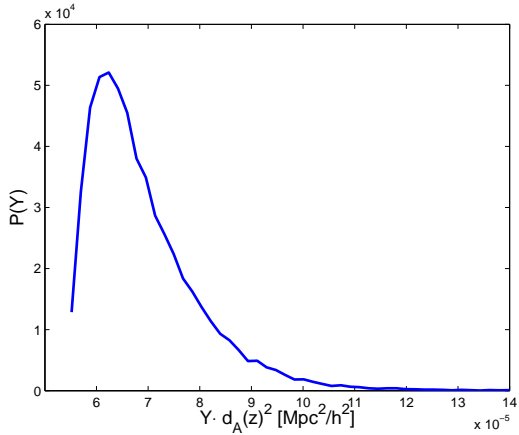, height=6cm}
\caption{PDF of the integrated Comptonization parameter from the merger-tree
model for $M=10^{15}h^{-1}M_{\odot}$ at $z=0.01$.}
\label{fig:Y_dist}
\end{figure}

Having determined the IC gas temperature and density profiles (as 
outlined above), we can now compute another key observable - the 
integrated Comptonization parameter. To do so, we also need to specify 
the gas mass fraction, which is taken to be $f_g=0.1$ for all halos. 
Figure \ref{fig:Y_dist} shows the PDF of $Y$; it exhibits the same 
general features as the temperature distribution, a sharp cutoff 
at low $Y$, and a long exponential tail at high values, largely due to 
clusters that formed uncharacteristically early. We note that a 
log-normal distribution is a poor fit to the outcome of our model. 

As in the case of IC gas temperature, scaling relations of the mean 
values of $Y$ can be computed. The scaling with mass is 
$Yd_A(z)^2\propto M^{\delta}$ where $d_A(z)$ is the angular diameter 
distance and $\delta$ varies with redshift, from $\delta=1.61$ for $z=0.01$ to
$\delta=1.64$ for $z=2$. This scaling is close to the standard result
$\delta=5/3$. Similarly, the Comptonization parameter scales with 
redshift as $Yd_A(z)^2\propto [E^2(z)\Delta_V(z)]^{\varepsilon}$ with 
$\varepsilon=0.26$ for $M=10^{14}h^{-1}M_{\odot}$ and $\varepsilon=0.31$ 
for $M=10^{16}h^{-1}M_{\odot}$. The evolution with redshift is slower 
than in the standard description where $\varepsilon=1/3$.

\subsection{Temperature number counts}

The PDFs of cluster observables presented above provide a theoretical 
basis for comparisons with results of cluster surveys. As an example we 
consider here the predicted temperature number 
counts, which is one of the statistical cluster functions that can be
used to determine cosmological parameters.

The temperature function, that is the cumulative number density of 
clusters above a certain temperature at a given redshift (interval) is 
computed from the following expression
\begin{equation}
	n(T_i)=\int_{M_{low}}^{M_{high}}{B(T_i|M,z)\frac{dn(M,z)}{dM}dM} , 
\end{equation}
where $dn(M,z)/dM$ is the 
mass function, namely the number
of halos per unit comoving volume per unit
mass. The selection function $B(T_i|M,z)$ is usually defined as $B=1$
if $T(M,z)>T_i$ and $B=0$ otherwise, where $T(M,z)$ is found according 
to the standard scaling relations (with a sharp cutoff).

However, in accord with our treatment here, there is no one-to-one 
correspondence between temperature and mass, so we need to incorporate 
the PDF of the temperature in the calculation of the number counts by 
using the following selection function
\begin{equation}
	B(T_i|M,z)=\int_{T_i}^{\infty}{P(T|M,z)dT} .
\end{equation}

\begin{figure}
\centering
\epsfig{file=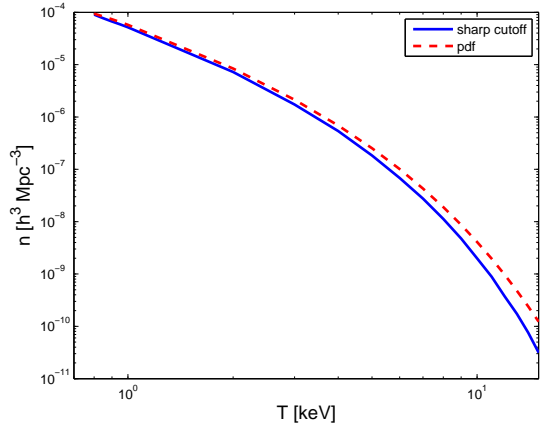, height=6cm}
\caption{Temperature number counts: the number density of clusters above a
certain temperature at $z=0.1$}
\label{fig:temp_nc}
\end{figure}

The temperature PDF is described by a log-normal distribution with expectation
value and variance taken from best fits to the results of our model. The
temperature functions calculated with our more realistic temperature PDF and
that with the standard relation (between temperature and mass) are shown in
Figure \ref{fig:temp_nc}. In the standard calculation we chose $T(M,z)$ to equal
the expectation value of the respective PDF. The calculations were performed
over the mass range $10^{13}-10^{16}h^{-1}M_{\odot}$ using the Sheth-Tormen mass
function.

The two calculations coincide for low temperatures, but for high 
temperatures our improved treatment yields appreciably higher
number counts. The reason for this is that our more exact treatment
takes into account the long tails of the distribution functions. 
Thus, low-mass clusters with mean temperatures below $T_0$, that do not 
contribute to $n(T_0)$ when the standard scaling is used, can have a 
significant overall contribution when the temperature PDF is used. As 
discussed earlier, this is due to the non-zero probability that the 
formation redshifts of these clusters, and hence also their 
temperatures, were higher than the mean values.

As we have mentioned earlier, the log-normal distribution is a mediocre fit to
the PDFs of high-mass clusters, and a better understanding of their shapes is
required in order to fully assess their impact on temperature number counts. The
above calculation demonstrates the importance of taking temperature PDFs into 
account in the analysis of cluster surveys.

\subsection{Model uncertainties}
\label{sec:model_unc}

In the previous sections we have shown that our method for the determination of 
the PDFs of the various cluster physical parameters provides a relatively 
simpler procedure to implement than hydrodynamical simulations. The procedure 
involves specifying several free parameters: $q$ - the maximal major merger
ratio, $\kappa$ - the parameter that determines the initial distance between
clusters, and the adiabatic index of the gas, $\Gamma$. 
We should also add to this list the parameters of the initial $c(M,z)$ chosen
for the smallest halos in the tree (see equation (\ref{eq:cfit_bullock})). These
parameters were found not to influence the results considerably when chosen
reasonably, in accordance with observational results and N-body simulations; see
the discussion at the end of Section \ref{sec:c_scaling}.

Since we are mainly interested in the global properties of the cluster, such as
the emission weighted temperature and the integrated Comptonization parameter,
our results have a very weak dependence on a particular choice for the IC gas
profile. We have repeated our calculations using the $\beta$-profile for the
gas:
\begin{equation}
	\rho(r)=\rho_{g0}\left[1+\left(\frac{r}{r_c} \right)^2 \right]^{-3\beta/2}
\end{equation}
with $\beta=2/3$. The gas core radius is given by $r_c=r_s/\eta$ where $r_s$ is
the DM scale radius, and a typical value is $\eta=2$ (Ricker and Sarazin,
2001). 
We have integrated the equation of hydrostatic equilibrium to obtain the gas 
temperature profile, setting the pressure to zero at infinity. With this profile, 
the temperature PDFs change by just a few percent relative to the polytropic 
model. We repeated the calculation using also the $\beta$-profile with a
different boundary condition, namely setting the gas temperature at the virial
radius to the temperature of the IGM, typically $10^6-10^7$ K. This too had only
a minor impact on the results.

\begin{figure}
\centering
\epsfig{file=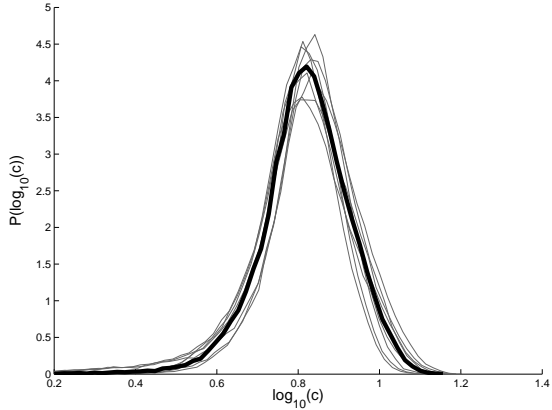, height=6cm}
\caption{Concentration parameter PDF from the merger-tree model for
$M=10^{15}h^{-1}M_{\odot}$ at $z=0$ and model parameters in the range
$q=7-13$ and $\kappa=4-6$. The thick curve shows the PDF calculated with the
fiducial values.}
\label{fig:cerr_all}
\end{figure}

In order to estimate the robustness of our model we compute the errors on the PDFs
that result from small deviations from the fiducial values of the main model
parameters. Figure \ref{fig:cerr_all} shows the PDF calculated with parameters in
the range $q=7-13,\: \kappa=4-6$.

Other relevant features of the PDF are the expectation value and its variance.
These quantities, especially for the temperature and Comptonization parameter,
need to be known in the analysis of cluster surveys. We thus need to determine
the uncertainty in their values due to variations of model parameters. Varying
the parameters in the range $q=7-13,\: \kappa=4-6,\: \Gamma=1.1-1.3$ we
obtain $<T>=9.9^{+0.9}_{-1.4}$ keV for $M=10^{15}h^{-1}M_{\odot}$ at
$z=0$, while the width of the distribution is $\sqrt{<(T-<T>)^2>}=1.4^{+0.2}_{-0.3}$ keV.
This likely is a conservative estimate for the range of the temperature
uncertainty.

The evolution of the variance of the PDF with mass and redshift and its uncertainty 
can also be estimated. For instance, if the parameters are varied in the same 
range $q=7-13,\: \kappa=4-6,\: \Gamma=1.1-1.3$, and for the same mass of 
$M=10^{15}h^{-1}M_{\odot}$ but for observation redshift of $z=0.5$, the following 
values are obtained: $<T>=12.0^{+1.0}_{-1.6}$ keV and $\sqrt{<(T-<T>)^2>}=1.6\pm 
0.3$ keV. The results for a mass of $M=10^{14}h^{-1}M_{\odot}$ at redshift $z=0$ 
are: $<T>=2.5^{+0.3}_{-0.4}$ keV and $\sqrt{<(T-<T>)^2>}=0.5^{+0.1}_{-0.09}$ keV. 
The relative uncertainties in all these cases are similar.

Finally, we can estimate the robustness of our results for the evolution 
of the scaling relations. As an example, we have checked how the scaling 
of the temperature with mass ($T\propto M^{\zeta}$) and redshift 
($T\propto[E^2(z)\Delta_V(z)]^{\lambda}$) changes when the model 
parameters are varied in the range $q=7-13,\: \kappa=4-6,\: 
\Gamma=1.1-1.3$. It turns out that $\zeta$ and $\lambda$ change by no 
more than $1\%$ and $6\%$, respectively, relative to the values obtained 
in Section \ref{sec:tem_scale}.

\section{Discussion}
\label{sec:discussion}

We have presented an expanded merger-tree treatment for the 
evolution of galaxy clusters that supplements the statistical
description of the dynamical evolution of DM halos with basic physical 
considerations that enable us to describe also the properties of IC gas.
It should be stressed again 
that our approach is statistical by construction and is not meant to provide 
a prescription for determining the structure of individual halos, but rather 
to serve as a tool for studying the properties of a population of clusters. 
While our treatment is essentially adiabatic, we have adopted an 
observationally-deduced value of the polytropic index. 
By doing so we partly compensate for the fact that gas cooling is not 
explicitly taken into account. Additional justification for the validity 
of our approach is the fact that we are interested here only in statistical 
properties of the cluster population, rather than in detailed spatial 
profiles of the gas density and temperature in individual (such as 
cooling-core) clusters.

We also assumed that the DM mass profile is not affected by
the IC gas. Although this approximation is often made in studies of the
statistical properties of a population of clusters (e.g. Bode, Ostriker \&
Vikhlinin 2009), it is likely to be inaccurate when radiative cooling is
important, or when there is energy exchange between the DM and the gas
components, for example during mergers. Numerical simulations (Duffy et al.
2010) show that there is a deviation of at most $15\%$ in the concentration
parameter of groups and clusters when baryonic physics is included, relative to
the DM only case. The impact of IC gas cooling on the DM density profile is
often described by adiabatic contraction models (e.g. Gnedin et al. 2004).
However, the assumption made in these models that the baryons initially trace
the DM distribution is violated during hierarchical build-up of halos. Indeed,
Duffy et al. (2010) found that results of the simulations were not well
described by adiabatic contraction models beyond $0.1R_{vir}$. A natural
extension of our model would be to incorporate IC gas in the halos that
constitute the merger tree and to follow the joint evolution of both components.

We calculated the PDFs of the cluster concentration parameter, its IC 
gas temperature, and integrated Comptonization parameter for different 
masses and redshifts of observation. Our deduced PDF of the concentration
parameter is well fit with a log-normal distribution, in accord with results
from N-body simulations. The temperature PDF for masses below 
$M\sim 2\cdot 10^{15}h^{-1}M_{\odot}$ can also be described with a log-normal
distribution. Our deduced mass-observable scaling relations are close to the
standard relations but contain some corrections - notably the evolution of
IC gas temperature with redshift is slower than in the simple model.
The results suggest that the gas temperature is not a separable function 
of mass and redshift. We show a possible application of our results to 
the analysis of cluster surveys by calculating IC gas temperature number 
counts, taking into account the effect of cluster formation history.

The probability density functions of the various observables can have 
important effects on the error estimation in the analysis of cluster 
X-ray and SZ surveys. As shown by Lima and Hu (2005), large 
uncertainties in the observable-mass distributions may substantially 
degrade the constraints on cosmological parameters from cluster surveys. 
The physically-based estimates of the PDFs of the observables considered 
here provide a tangible basis to begin addressing this aspect. 

Among the other related applications of the approach presented here a 
particularly timely one is the calculation of the SZ power spectrum, 
which will be mapped by the {\it Planck} satellite and several 
ground-based SZ projects. Comparisons of results from our merger-tree 
approach and those from simulations and semi-analytical treatments 
(e.g., see Sadeh, Rephaeli \& Silk 2007 and references therein) will
yield important insight that will help gauging the relative merits 
and disadvantages of these very different approaches.

\section*{Aknowledgements}
The authors wish to thank the GALFORM team for making the code publicly
available. 
This research was supported by a US-Israel Binational Science Foundation 
grant 2008452.

\bsp

\label{lastpage}

\end{document}